%% file: main.tex
\documentclass[10pt,twocolumn]{article} 
\usepackage{arxivStyle}
\usepackage{times}
\usepackage{graphicx}
\usepackage{amsmath}
\usepackage{amssymb}
\usepackage{url,hyperref}
\usepackage{pifont}

\begin{document}

\title{Automatic Skull Reconstruction by Deep Learnable Symmetry Enforcement}

\author{Marek Wodzinski$^{1,2}$, Mateusz Daniol$^{1}$, Daria Hemmerling$^{1}$ \\
\\
$^{1}$Department of Measurement and Electronics \\
AGH University of Kraków, Kraków, Poland \\
wodzinski@agh.edu.pl  \\
$^{2}$ University of Applied Sciences Western Switzerland (HES-SO Valais) \\
Institute of Informatics, Sierre, Switzerland \\
}

\maketitle
\thispagestyle{empty}

\begin{abstract}
\textbf{Background and objective:} Every year, thousands of people suffer from skull damage and require personalized implants to fill the cranial cavity. Unfortunately, the waiting time for reconstruction surgery can extend to several weeks or even months, especially in less developed countries. One factor contributing to the extended waiting period is the intricate process of personalized implant modeling. Currently, the preparation of these implants by experienced biomechanical experts is both costly and time-consuming. Recent advances in artificial intelligence, especially in deep learning, offer promising potential for automating the process. However, deep learning-based cranial reconstruction faces several challenges: (i) the limited size of training datasets, (ii) the high resolution of the volumetric data, and (iii) significant data heterogeneity.

\textbf{Methods:} In this work, we propose a novel approach to address these challenges by enhancing the reconstruction through learnable symmetry enforcement. We demonstrate that it is possible to train a neural network dedicated to calculating skull symmetry, which can be utilized either as an additional objective function during training or as a post-reconstruction objective during the refinement step. We quantitatively evaluate the proposed method using open SkullBreak and SkullFix datasets, and qualitatively using real clinical cases.

\textbf{Results:} The results indicate that the symmetry-preserving reconstruction network achieves considerably better outcomes compared to the baseline (0.94/0.94/1.31 vs 0.84/0.76/2.43 in terms of DSC, bDSC, and HD95). Moreover, the results are comparable to the best-performing methods while requiring significantly fewer computational resources ($<$ 500 vs $>$ 100,000 GPU hours). Moreover, its relatively low computational complexity makes it scalable for reconstructing all symmetrical structures.

\textbf{Conclusions:} The article introduces an automatic skull reconstruction method based on the enforcement of skull symmetry using a learnable deep learning network. The method requires significantly fewer computational resources compared to other well-performing methods and is able to improve the reconstruction for the out-of-distribution cases. The proposed method is a considerable contribution to the field of applied artificial intelligence in medicine and is a step toward automatic cranial defect reconstruction in clinical practice.

\end{abstract}

\section{Introduction}

The automatic reconstruction of cranial defects is an important research area. Every year thousands of people suffer from cranial damage due to traffic accidents, warfare, or surgery~\cite{samii2002skull,de2012cranial}. The following skull reconstruction is a costly and time-consuming process~\cite{bonda2015recent}, resulting in a large group of people being excluded from professional and social life. One of the factors that increase the waiting time for the reconstruction is connected with modeling the personalized implants.
 
The cranial defects are highly heterogeneous, with different sizes, shapes, and thicknesses, resulting in the necessity to model and manufacture personalized solutions~\cite{fishman2024thickness}. Each implant is unique - no two skulls are the same, and neither are the defects. As a result, the current state-of-the-art in personalized implant reconstruction is based on computer-aided design (CAD) software used by experienced biomechanics, resulting in costly and time-consuming procedure~\cite{ameen2018design,marreiros2016custom}.

Researchers are attempting to speed up the modeling process by using deep neural networks. The idea was greatly accelerated by two editions of the AutoImplant challenge organized during the MICCAI conference~\cite{li2021autoimplant,li2023towards} where authors proposed numerous successful contributions~\cite{li2020baseline,kodym2020cranial,kodym2021deep,ellis2020deep,matzkin2020cranial,wodzinski2021improving,jin2020high,pathak2021cranial,mahdi2021u,li2021learning,bayat2020cranial,kroviakov2021sparse,wodzinski2022deep}. Afterwards, researchers continued working on the topic by proposing occupancy networks~\cite{mazzocchetti2024automatic}, exploring sparse solutions~\cite{li2023sparse}, using the most recent advances in deep architectures~\cite{kesornsri2024cranext}, or reformulating the problem into a point cloud completion task~\cite{friedrich2023point,wodzinski2023high,sulakhe2022crangan}. The most recent and successful contribution is based on strong data augmentation to increase the dataset heterogeneity, consisting of image registration and latent diffusion models~\cite{wodzinski2024improving}. Even though the approach is successful, resulting in great generalizability and almost perfect reconstructions for real defects, its limitation is connected with the enormous computational resources required, limiting its applicability to different organs and types of damages. 

Automatic skull reconstruction is connected with several challenges: (i) limited training dataset size, (ii) high volumetric data resolution, and related VRAM issues, and (iii) enormous data heterogeneity. Almost all current solutions are trained using synthetic defects because it is close to impossible to acquire ground-truth real defects. It requires computed tomography (CT) volumes acquired before and after e.g. traffic accident which is obviously undesirable. Although high volume resolutions can be solved by recent advances in hardware and sparse deep networks~\cite{li2023sparse}, the limited amount of data and its heterogeneity remain unsolved tasks that all involved researchers attempt to solve.

\begin{figure}[!htb]
    \centering
    \includegraphics[width = 0.45\textwidth]{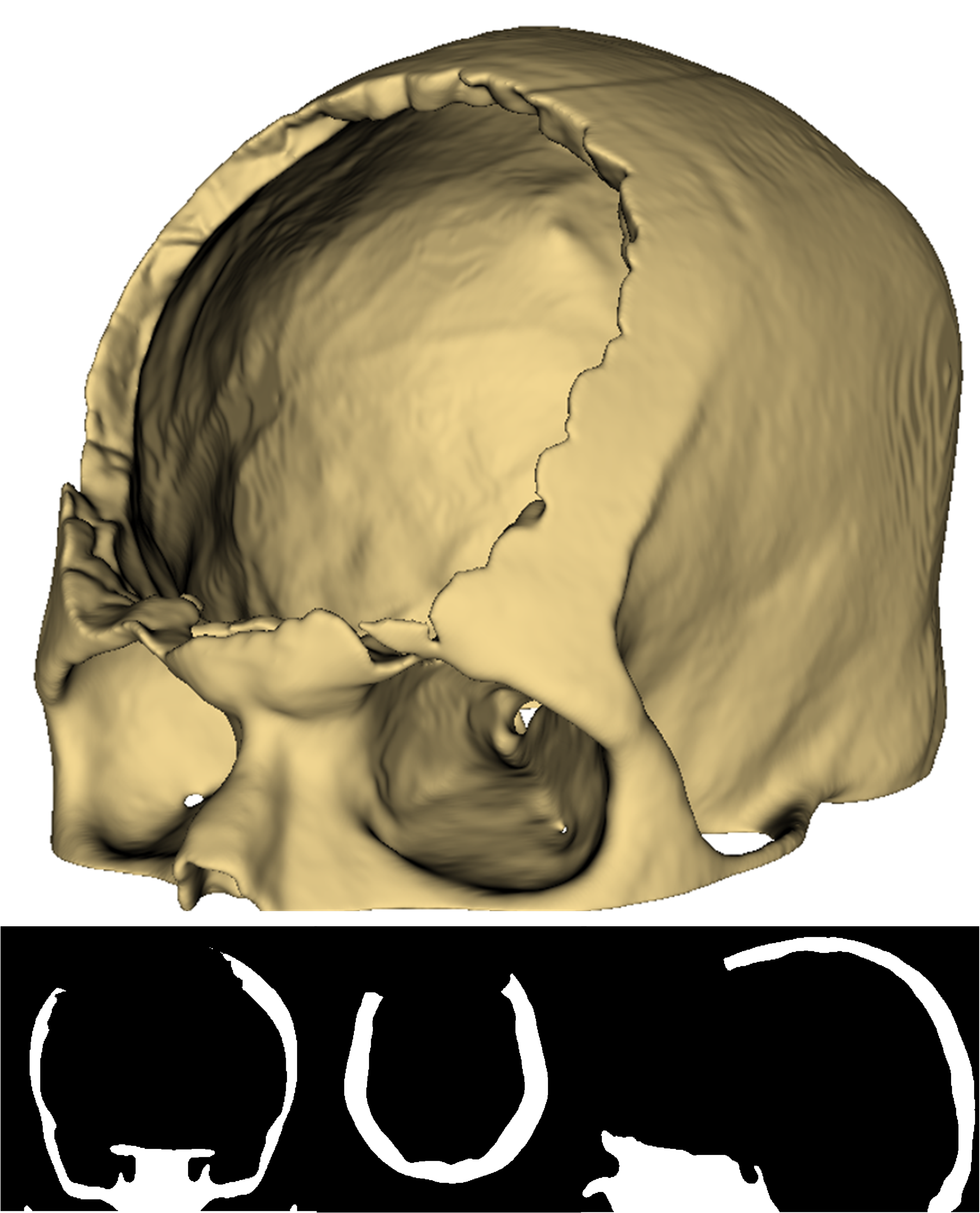}
    \caption{Example of a damage in the frontal bone that cannot be reconstructed directly by reflecting about the symmetry axis because the defect breaks the skull symmetry resulting in missing data.}
    \label{fig:limitation}
\end{figure}

There exists an interesting, yet unexplored, field of research in the context of skull reconstruction connected with learnable symmetry calculation~\cite{gao2020prs, zhou2021nerd}. The idea is to calculate the symmetry axis of a given object by using deep neural networks. 
We argue that it can also be used to improve the automatic reconstruction of cranial defects. Although the skulls are not perfectly symmetric, especially in the craniofacial regions, the bones forming the brain skull are close to being symmetric~\cite{ratajczak2021symmetry}. Moreover, symmetry is usually desired from an aesthetic perspective. This observation can be used to improve automatic reconstruction algorithms. There are initial works that attempt to reconstruct the defects by mirroring~\cite{senck2013virtual,chauhan2022reconstruction}, however, they are limited by the necessity of manually defining the flip axis. Moreover, they cannot be applied if the skull damage goes through the symmetry axis (e.g., in the frontoorbital skull) due to missing data (see Figure~\ref{fig:limitation}). To our knowledge, no contributions combine both the learnable symmetry axis calculation with the personalized defect reconstruction. Nevertheless, as Herman Weyl states: "Symmetry expresses not only the aesthetic side of mathematics but also its power to predict, explain, and understand the laws of nature."~\cite{weyl1938symmetry} and we confirm the statement in this work.

\textbf{Contribution: } In this work, we propose a novel method dedicated to efficient and fast personalized reconstruction of cranial defects by enforcing the symmetry of the reconstructed skull. We propose a method based on learnable symmetry using deep vision transformers that can be integrated into both segmentation-based reconstruction and registration-based refinement. We show that the proposed method is on par with the best-performing reconstruction methods, however, requires significantly fewer computational resources. Moreover, registration-based refinement can be considered as a form of instance optimization that improves the quality of the reconstruction for out-of-distribution cases. The proposed method is a huge step towards using deep learning-based reconstruction methods in everyday clinical practice.

\section{Methods}

\subsection{Overview}

\begin{figure*}[!htb]
    \centering
    \includegraphics[width = 0.99\textwidth]{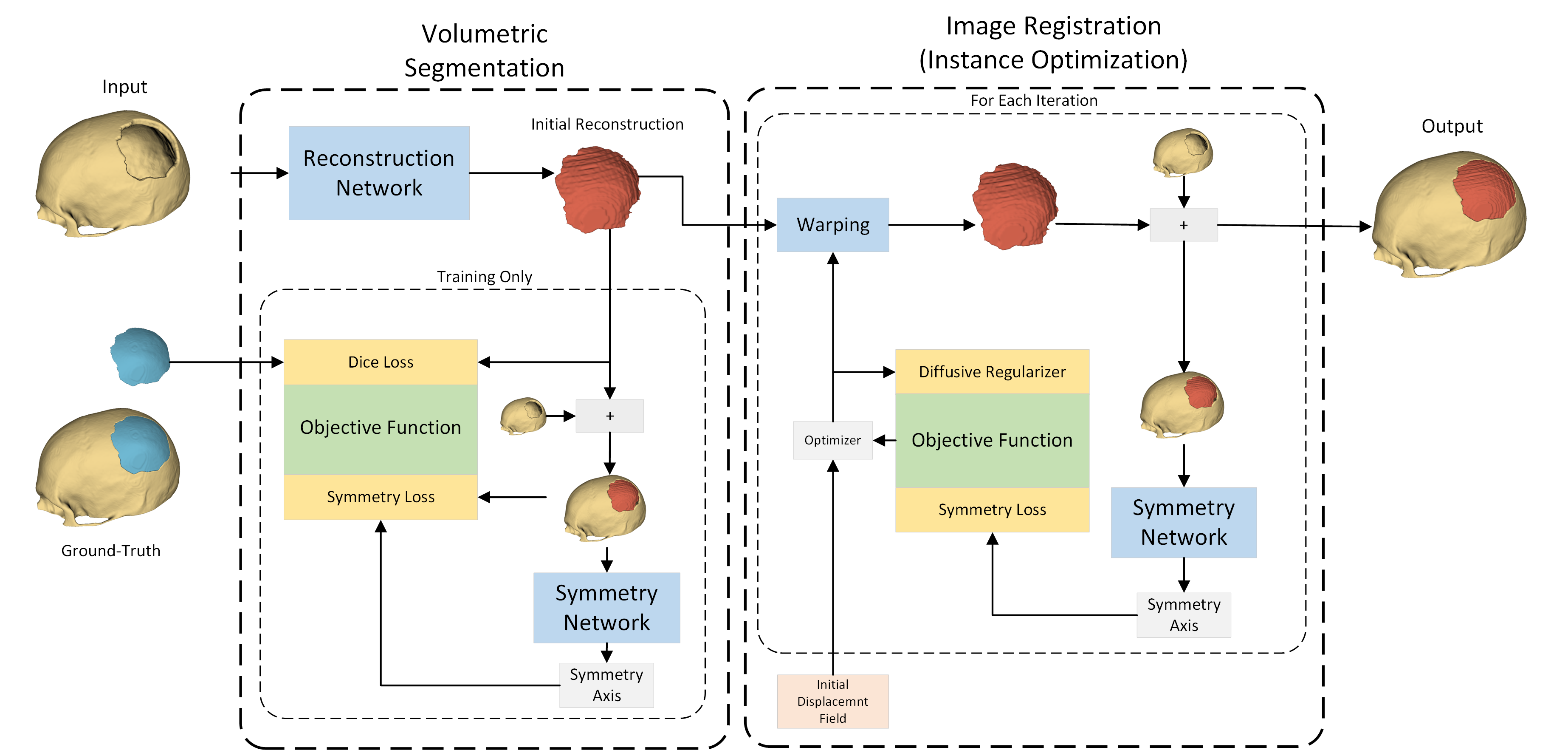}
    \caption{An overview of the proposed method. The method starts with the segmentation-based volumetric reconstruction enhanced by the proposed symmetry network and loss. In the second step, the initial reconstruction is fine-tuned by iterative image registration using the symmetry loss as the objective function.}
    \label{fig:pipeline}
\end{figure*}

The proposed method consists of two cooperating deep neural networks. The first one is a classical encoder-decoder architecture dedicated to volumetric segmentation, in this work responsible for the defect reconstruction, further called a "reconstruction network" (RN). The second one is an encoder-regressor architecture responsible for calculating the axis of symmetry of a healthy skull, further denoted as the "symmetry network" (SN). We use the SN to extend the deep learning-based volumetric reconstruction in two possible ways: (i) to enforce symmetry after the reconstruction by using it as an additional objective function during training the RN, and (ii) as an optional post-reconstruction refinement step during deformable image registration using the SN as an objective function. The overview of the processing pipeline is shown in Figure~\ref{fig:pipeline}.

\subsection{Symmetry network and symmetry loss}

The symmetry network (SN) takes as input a healthy skull and is responsible for calculating the plane equation being the symmetry axis. The SN is an encoder-regressor architecture, where the encoder is responsible for feature extraction and the regressor calculates the axis of symmetry. In this work, we evaluated two encoder backbones based on volumetric ResNet and Vision Transformer (ViT) to verify whether the inductive bias and receptive field of the encoder architecture are important factors for calculating the axis of symmetry. The training was unsupervised, we introduced a new objective function named Symmetry Loss (SL), defined as:
\begin{equation}
    \begin{split}
    \pi &= SN(V) \\
    SL(V, \pi) &= DSL(V, R(V, \pi)),
    \end{split}
\end{equation}
where $SN$ is the symmetry network, $DSL$ is the Dice Loss, $V$ is the input volume (healthy skull), $R$ is the reflection operator, $\pi$ is the calculated plane equation. The operation is performed by direct coordinate reflection followed by bilinear interpolation (nearest-neighbor interpolation would be non-differentiable). During training, healthy skulls were augmented by random flipping across all axes, and random affine transformations, including rotation, translation, and scaling, without random shear.

\subsection{Symmetry in segmentation}

The SL can be used as an additional objective function during the RN training to enforce symmetry of the reconstructed skull (Figure~\ref{fig:pipeline}). The training of baseline RN involves supervised cost function dedicated to volumetric segmentation, in this work the DSL. The procedure for using the SL during the RN training is slightly different compared to the SN training:
\begin{enumerate}
    \item Forward the defective skull through the RN and obtain initial reconstruction $Rec$.
    \item Combine the initial reconstruction with the defective skull $V_{Rec}$.
    \item Forward the result through the SN network to calculate the axis of symmetry $\pi$.
    \item Reflect the reconstructed defect about the calculated axis of symmetry and calculate the SL with respect to the reconstructed skull.
\end{enumerate}
Therefore, the final objective function becomes:
\begin{equation}
    \begin{split}
        Rec &= RN(V) \\
        V_{Rec} &= V \vee Rec \\
        \pi &= SN(V_{Rec}) \\
        O_{Rec}(V, Gt) &= DSL(Rec, Gt) + \alpha SL(V_{Rec}, \pi),
    \end{split}
\end{equation}
where $V_{Rec}$ is the reconstructed skull, $Gt$ is the ground-truth defect, and $\alpha$ controls the influence of the SL on the final reconstruction. The procedure is fully differentiable. Since both DL and SL are strongly correlated and both are within the same range of values [0-1], we set $\alpha$ to 1 in all experiments.

\subsection{Symmetry in registration}

The SL can also be used as an objective function after reconstruction, during the optional refinement phase (Figure~\ref{fig:pipeline}). The refinement phase is formulated as a deformable registration based on instance optimization, with a highly regularized displacement field:
\begin{equation}
    \begin{split}
    Rec &= RN(V) \\
    Rec &= Rec \circ u \\
    V_{Rec} &= V \vee Rec \\
    \pi &= SN(V_{Rec}) \\
    O_{Reg}(V, u) &= SL(V_{Rec}, \pi) + \lambda Reg(u),
    \end{split}
\end{equation}
where $O_{Reg}$ is the objective of the registration, $SL$ is the symmetry loss, $u$ denotes the dense displacement field, $Reg$ is the diffusive regularization, $\circ$ is the warping operator, and $\lambda$ controls the smoothness of the displacement field. The procedure for calculating $SL$ is the same as during the RN training.

The procedure aims to deform the reconstructed defect in a way that makes the resulting healthy skull as symmetrical as possible. Importantly, only the reconstructed defect is deformed, the defective skull is not undergoing any deformable transformations. The procedure slightly increases the reconstruction time; however, it enables the user to handle out-of-distribution cases for which the RN does not generalize well. If the reconstructed implant is at least partially correct, it means that it at least partially fills the cranial cavity with reasonable size and shape; it can be geometrically transformed into a solution that preserves skull symmetry.

\subsection{Dataset}

We use open SkullBreak and SkullFix datasets to train and evaluate the proposed method~\cite{kodym2021skullbreak}. The training subset of the SkullBreak dataset consists of 114 skulls, each with 5 different defects that are inserted randomly in the frontoorbital, bilateral, parietotemporal bones, resulting in 570 training cases. The test subset of the SkullBreak dataset consists of 100 defective cases. In contrast, SkullFix consists of 100 defective skulls in the training subset and 110 cases in the testing subset. Both the datasets include real skulls with synthetic defects. The reason why the defects are synthetic is connected with the difficulty of acquiring ground-truth annotations; it would require to get CT volumes before and after the traumatic event.

\begin{figure}[!htb]
    \centering
    \includegraphics[width = 0.45\textwidth]{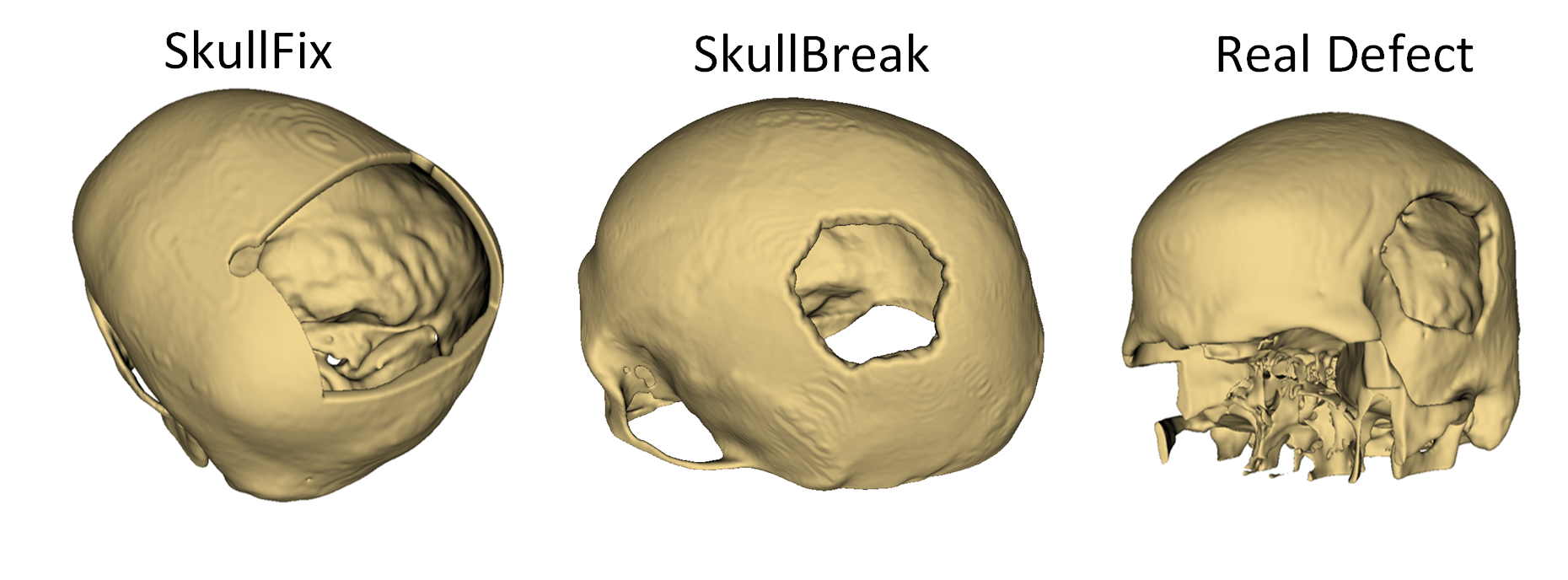}
    \caption{Exemplary cases from the three datasets: (i) the SkullFix containing real skulls with synthetic defects in parietal bones, (ii) the SkullBreak consisting of real skulls with synthetic defects in all areas of skull, (iii) the real skulls with real, clinical defects.}
    \label{fig:dataset}
\end{figure}

In addition, we perform a qualitative evaluation using 11 real clinical cases that were released to evaluate generalizability subtasks in the AutoImplant challenge. In contrast to the SkullFix and SkullBreak datasets, these cases are real skulls with real cranial damage. Unfortunately, because ground-truth does not exist for such cases, we cannot perform a quantitative comparison. Visualization presenting exemplary cases from these three datasets is shown in Figure~\ref{fig:dataset}.

\subsection{Experimental Setup}

We performed several ablation studies to verify the impact of the proposed method. First, we calculated the maximum quantitative performance that could be achieved by the learnable symmetry enforcement and compared two architectures implementing the SN. Then, we compared several architectures for RNs, and their impact for the automatic cranial defect reconstruction. We continued by comparing the segmentation-based approach with the registration-based refinement. Finally, we compared the proposed method to the state-of-the-art methods.

All training of SNs and RNs were performed until convergence using the AdamW optimizer with an initial adaptable learning rate equal to 0.001. We did not include any folding or ensembling mechanism, all the experiments were performed using the same 90/10 random validation split. The checkpoint with the best performance on the validation subset were used to perform the evaluation on the test subset. The training was performed using PLGrid Athena supercomputer containing computing nodes with NVIDIA A100 GPUs. Inference was performed using a workstation equipped with NVIDIA A6000 GPUs. The reconstruction, without the registration-based refinement, did not require more than 12GB of dedicated VRAM memory. The registration-based refinement increased the VRAM usage to 18 GBs. All operations (pre-processing, reconstruction, refinement, post-processing) were accelerated using GPU. 

Quantitative comparisons are performed using the Dice Score (DSC), 95th percentile of Hausdorff distance (HD95), surface Dice Score (SDSC), and mean surface distance (MSD). Due to missing ground-truth, these metrics cannot be calculated for the real cases, however, from the qualitative comparison it is clear that the symmetry enforcement improves the quality of reconstruction. Any claim in the Results and Discussion sections about a statistically significant improvement is supported by a Wilcoxon signed-rak test with p-value below 0.01. The quantitative comparisons are performed using the test subsets of the SkullFix and SkullBreak datasets. The test sets were not used at all during the training of SNs or RNs.

\section{Results}

\subsection{Symmetry limitations and architectures}

The first step to evaluate the potential of the symmetry enforcement was to calculate the best possible outcomes by SN. To do that, we applied the SN to all complete skulls and associated implants from SkullBreak and SkullFix datasets and calculated the DSC between the implant reflected about the symmetry axis and the overlapping part of the skull. We compared two encoders, the first based on a volumetric ResNet and the second based on a volumetric vision transformer (ViT). The results are presented in Table~\ref{tab:sn_limit}. The results confirm that the ViT-based network was significantly better than the ResNet-based encoder. 

The experiment achieved results (avgDSC = 0.956/0.969 for SkullBreak and SkullFix, respectively) outperforming most state-of-the-art reconstruction methods and confirmed that the idea is promising. However, in this experiment, the generalizability of the network was not important. The idea behind it was connected with verifying the high-level idea. To verify its usability in practice, the symmetry enforcement had to be integrated with the defect reconstruction algorithm.

\input{tables/table_symmetry_limitations}

\subsection{Reconstruction architecture}

The next step was to evaluate the impact of the RN architecture on the defect reconstruction and the influence of SN. We compared several architectures: (i) RUNet, (ii) SwinUNETR, (iii) UNETR, (iv) AttentionUNet, and (v) SegResNet - with and without the additional objective function based on enforcing the skull symmetry. The results are presented in Table~\ref{tab:architectures}. The results confirm that the inductive bias introduced by CNNs is beneficial for defect reconstruction and that transformer-based networks do not improve the reconstruction. On the other hand, symmetry enforcement improves the results for all the architectures.

\input{tables/table_architectures}

\subsection{How to use the symmetry enforcement?}

The results presented in Table~\ref{tab:architectures} confirmed that symmetry enforcement during RN training improves the results. However, the results were not as good as the best possible outcomes presented in Table~\ref{tab:sn_limit}. The reason why the results are suboptimal is related to the generalizability of the network. Although we enforced the network to improve the symmetry of the reconstructions, the procedure did not ensure that the reconstruction network generalizes to previously unseen cases or distributions.

\begin{figure*}[!htb]
    \centering
    \includegraphics[width = 0.80\textwidth]{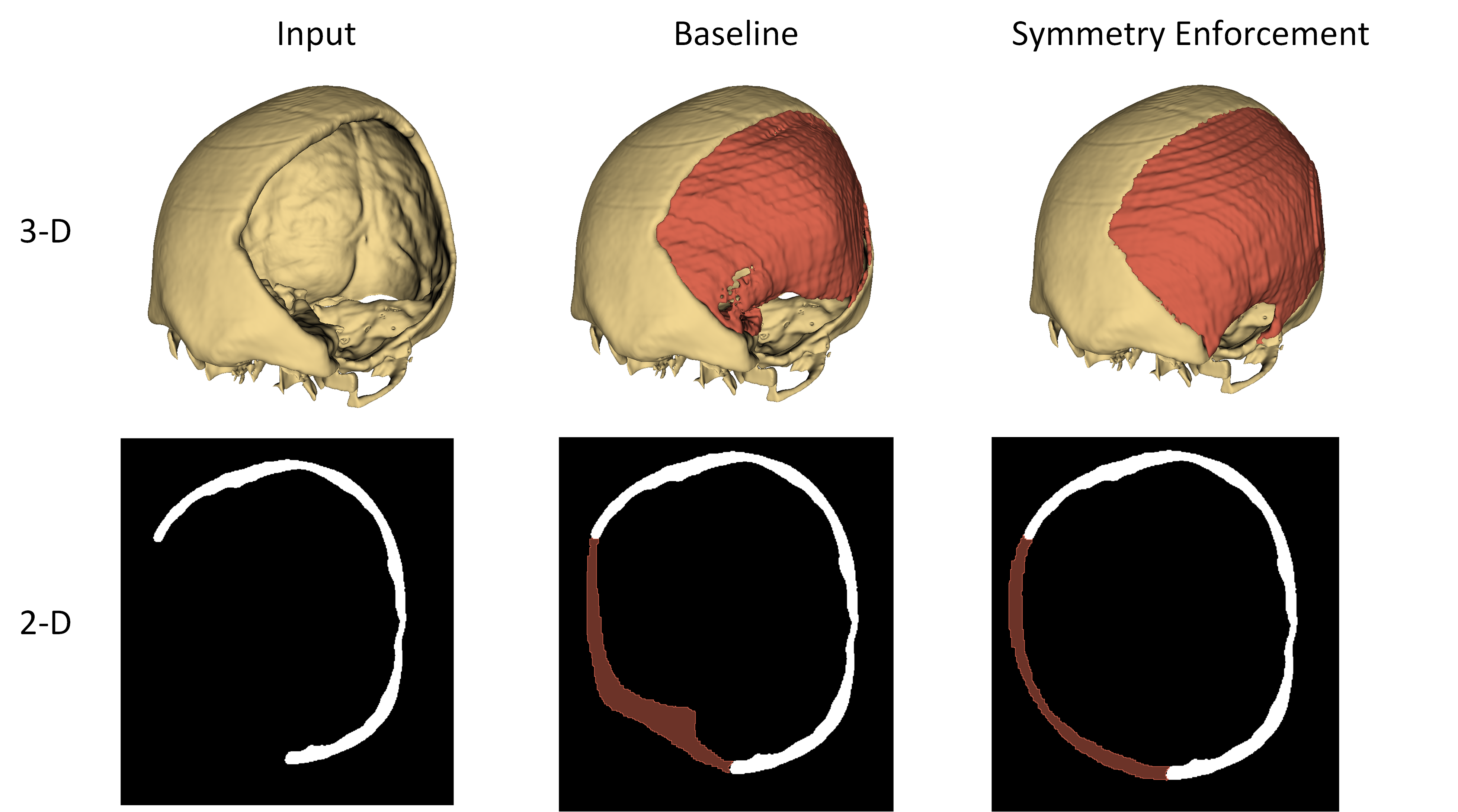}
    \caption{The influence of the symmetry enforcement on a real defective case. Note that the baseline is unable to generalize into a large real defect. Nevertheless, the symmetry enforcement considerably improves the reconstruction quality, leaving only a small hole between the frontal and parietal bones (the image registration-based refinement limited by the diffusive regularization could not deform the initial reconstruction further).}
    \label{fig:example}
\end{figure*}

Therefore, we decided to reformulate the problem and enforce the symmetry during an optional refinement step based on the image registration, performed after the reconstruction. The results comparing segmentation- and registration-based symmetry enforcement are presented in Table~\ref{tab:seg_vs_reg}. Exemplary qualitative results that present the influence of symmetry enforcement are shown in Figure~\ref{fig:example}. The results of the registration-based refinement and close the best-possible ones (Table~\ref{tab:sn_limit}). Moreover, the post-reconstruction refinement step is an instance optimization procedure, allowing one to improve the results for out-of-distribution cases for which the initial reconstruction is only partially correct.

\input{tables/table_seg_vs_reg}

\subsection{State-of-the-art \& Real clinical cases}

Finally, we compared the proposed method to other state-of-the-art methods (Table~\ref{tab:sota}). The proposed method outperformed all methods, except for the contribution that uses strong data augmentation by image registration and latent diffusion models. Moreover, the proposed method successfully reconstructed all 11 clinical cases, leaving only small holes in the largest defects where the registration-based refinement was limited by the strong diffusive regularization, compared to only 6 accurate reconstructions by the baseline without the proposed symmetry enforcement.

\input{tables/table_sota}

\section{Discussion}

The proposed method outperforms most of the state-of-the-art solutions (Table~\ref{tab:sota}). It outperforms all point cloud-based methods and almost all volumetric approaches. The quantitative results are worse only when compared to heavily augmented reconstruction methods requiring enormous computational resources. The DSC, bDSC, and HD95 are 0.94/0.95, 0.94/0.95, 1.31/1.28 respectively (SkullBreak/SkullFix) and are significantly improved over the baseline (0.84/0.89, 0.76/0.88, 2.43/1.88). Such results can be considered acceptable from the clinical point of view and are ready to be fine-tuned into personalized implants.

As shown in Table~\ref{tab:seg_vs_reg}, the symmetry enforcement may be used both as an additional objective function during training, as well as an optional refinement step after reconstruction by using it as an objective for the following image registration. Both approaches have advantages and limitations. Using it as an additional objective function during training results in faster inference (average  $<$ 5 seconds), however, it does not guarantee that the reconstruction method will generalize correctly to previously unseen cases. On the other hand, the registration-based refinement is slower (average  $<$ 15 seconds), however, it can improve reconstruction for out-of-distribution cases that were incorrectly reconstructed during the first step. The SN has great generalizability; therefore, using it as an objective function will lead to the correct solution, as long as the initial reconstruction is at least approximately good (the regularization term would not allow too large displacements or significant changes in volume). Moreover, any time below several minutes is acceptable from the clinical point of view; the manufacturing and sterilization processes require more time.

An interesting observation comes from Table~\ref{tab:architectures}. It shows that the RUNet-based reconstruction architecture outperforms SwinUNETR and other transformer-based methods. On the other hand, the ViT outperforms the convolutional encoder in the symmetry calculation (Table~\ref{tab:sn_limit}). It shows that the large receptive field of ViT is beneficial for symmetry calculation; however, it is not important to the reconstruction itself where the inductive bias introduced by CNNs may even be beneficial. The best performing setup is therefore a combination of ViT for the skull symmetry calculation and RUNet for the defect reconstruction.

The proposed method has considerable advantages compared to the strongly augmented reconstruction methods (Table~\ref{tab:sota}). Although the quantitative results are insignificantly lower, the method requires significantly fewer computational resources during training, the same computational resources during segmentation-based inference, and an acceptable (and optional) increase in reconstruction time by using registration-based refinement. The segmentation-based reconstruction pipeline requires less than 5 seconds for the full reconstruction (pre-processing, reconstruction, post-processing) while adding the registration-based refinement increases the time to less than 15 seconds. Therefore, both versions can be considered real-time methods that could potentially be used directly during surgical interventions. Training both the symmetry preserving and the reconstruction networks required less than 500 A100 GPU hours, compared to more than 100,000 hours required by the methods supported by the generative augmentation~\cite{wodzinski2024improving}. As a result, the proposed method is more scalable and can be applied to other symmetric structures with relatively low computational costs. 

Another advantage of the proposed method is related to its ability to reconstruct large defects. The CNN-based reconstruction networks work well for small defects; however, large reconstructions are highly problematic, probably due to the limited receptive field. Thanks to post-reconstruction refinement, the proposed method allows the user to improve large reconstructions and make them more accurate (Figure~\ref{fig:example}).

However, the proposed method has several limitations. One of them is related to the fact that the algorithm reconstructs the defect but does not model the personalized implant. Personalized implants are not the same as reconstructed defects. Custom implants may require different thicknesses, shapes, or rims to be implantable~\cite{fishman2024thickness}.

Another limitation, although only apparent, is that there are worse quantitative results for some cases. Enforcing the symmetry of the reconstructed skull can sometimes decrease the quantitative results. However, one should remember that the real skulls are not perfectly symmetrical~\cite{ratajczak2021symmetry}, but the symmetry is actually desirable~\cite{rhodes2002facial,penton2001symmetry}. People with symmetric faces and heads are considered more attractive~\cite{grammer1994human}, leading to many plastic surgeries.

Finally, if the initial reconstruction is completely incorrect, the registration-based refinement will not be able to improve the reconstruction. The diffusive regularization does not allow large deformable displacement of the defect, neither enables the creation of synthetic data. Probably combining the SL with diffusion sources could lead to perfect reconstructions even for such cases; however, it is a topic for further investigation.

The proposed method is of significant clinical importance. Every year, thousands of people suffer from cranial damage and require personalized reconstructions. The ability to model personalized implants automatically and in real-time has significant practical implications. For example, it would allow for the closure of the cranial cavity directly during primary surgery, without the need to plan the reconstruction surgery. Although it is impossible for some surgeries (e.g. due to brain swelling), preventing even some of them would be a great achievement. Nevertheless, reducing the waiting time and costs of the procedure for other patients is equally important.

In future work, we plan to propose a method dedicated to transferring the reconstructed defects to personalized implants. Reconstructed defects cannot be used directly for 3-D printing or other manufacturing methods because they must be adapted to the desired implant rim and thickness, material, and shape of the outer cavity~\cite{fishman2024thickness}. Therefore, a method dedicated to automatically modifying the reconstructed defects based on the desired properties is required.

To conclude, we proposed a novel cranial defect reconstruction method based on enforcing the symmetry of the reconstructed defects. The proposed method significantly improves the quality of the reconstruction compared to the baselines and is as accurate as the best-performing methods while requiring significantly fewer computational resources. Moreover, since most of the rigid structures in the human body are approximately symmetric, the approach can be extended to reconstruct different symmetric organs and bones. The work is a considerable contribution to the field of artificial intelligence in medicine.

\section*{Acknowledgments}

The project was funded by The National Centre for Research and Development, Poland under Lider Grant no: LIDER13/0038/2022 (DeepImplant). We gratefully acknowledge Polish HPC infrastructure PLGrid support within computational grants no. PLG/2024/017079.

\section*{Ethical Approval}

All data used in the study were acquired retrospectively and acquired respective approvals~\cite{kodym2021skullbreak}. The study was approved by the Bioethics Committee of the AGH University of Krakow. 

\section*{Disclosure of Interests} The authors have no competing interests to declare that are relevant to the content of this article.

\bibliographystyle{abbrv}
\bibliography{main}
\end{document}

%% file: tables/table_symmetry_limitations.tex
\begin{table*}[!htb]
\centering
\caption{Table presenting the best achievable outcomes by the Symmetry Network (SN) after reflecting the healthy skull about the calculated symmetry axis and evaluating the DSC between the ground-truth defect and the corresponding part of the reflected skull. Note that the SN generalizes very well - the results on the test subset follow the results on the training and validation subsets, however, the ResNet-based solutions is unable to converge to accurate reflections.}
\renewcommand{\arraystretch}{1.0}
\footnotesize
\resizebox{0.60\textwidth}{!}{%
\begin{tabular}{lccccccc}
\label{tab:sn_limit}
Network & \multicolumn{3}{c}{SkullBreak} & \multicolumn{3}{c}{SkullFix} \tabularnewline
\hline
\multicolumn{1}{c}{} & \multicolumn{1}{c}{Train} & \multicolumn{1}{c}{Validation}  & \multicolumn{1}{c}{Test} & \multicolumn{1}{c}{Train} & \multicolumn{1}{c}{Validation} & \multicolumn{1}{c}{Test} \tabularnewline

ResNet & 0.903 & 0.892 & 0.907 & 0.911 & 0.905 & 0.896
\tabularnewline
ViT & 0.942 & 0.958 & 0.956 & 0.973 & 0.952 & 0.969
\tabularnewline
\hline
\end{tabular}}

\end{table*}

%% file: tables/table_architectures.tex
\begin{table*}[!htb]
\centering
\caption{Table presenting the comparison of different Reconstruction Network (RN) architectures and the influence of using the Symmetry Network (SN). Note that the traditional RUNet architecture outperforms the transformer-based variants for the RN, however, the SN improves the results for all the networks.}
\renewcommand{\arraystretch}{1.0}
\footnotesize
\resizebox{0.99\textwidth}{!}{%
\begin{tabular}{lcccccccc}
\label{tab:architectures}
Network & \multicolumn{4}{c}{SkullBreak} & \multicolumn{4}{c}{SkullFix} \tabularnewline
\hline
\multicolumn{1}{c}{} & \multicolumn{1}{c}{DSC $\uparrow$} & \multicolumn{1}{c}{SDSC $\uparrow$}  & \multicolumn{1}{c}{HD95 [mm] $\downarrow$} & \multicolumn{1}{c}{MSD [mm] $\downarrow$} & \multicolumn{1}{c}{DSC $\uparrow$} & \multicolumn{1}{c}{SDSC $\uparrow$}  & \multicolumn{1}{c}{HD95 [mm] $\downarrow$} & \multicolumn{1}{c}{MSD [mm] $\downarrow$}
\tabularnewline
\hline
\multicolumn{9}{c}{Without Symmetry Enforcement ($\alpha=0$)}
\tabularnewline
\hline
RUNet & 0.843 & 0.743 & 2.438 & 0.787 & 0.892 & 0.881 & 1.882 & 0.554
\tabularnewline
SwinUNETR & 0.702 & 0.658 & 3.214 & 1.092 & 0.785 & 0.792 & 2.818 & 0.721
\tabularnewline
UNETR & 0.670 & 0.632 & 4.013 & 1.187 & 0.779 & 0.797 & 2.884 & 0.725
\tabularnewline
AttentionUNet & 0.741 & 0.708 & 3.170 & 0.887 & 0.841 & 0.857 & 2.328 & 0.667
\tabularnewline
SegResNet & 0.729 & 0.702 & 3.279 & 0.952 & 0.819 & 0.832 & 2.452 & 0.681
\tabularnewline
\hline
\multicolumn{9}{c}{With Symmetry Enforcement ($\alpha=1$)}
\tabularnewline
\hline
RUNet & 0.904 & 0.902 & 1.678 & 0.539 & 0.927 & 0.923 & 1.437 & 0.389
\tabularnewline
SwinUNETR & 0.791 & 0.739 & 2.517 & 0.798 & 0.866 & 0.872 & 2.128 & 0.641
\tabularnewline
UNETR & 0.768 & 0.735 & 2.618 & 0.801 & 0.864 & 0.874 & 2.124 & 0.639
\tabularnewline
AttentionUNet & 0.829 & 0.788 & 2.271 & 0.761 & 0.896 & 0.892 & 1.842 & 0.542
\tabularnewline
SegResNet & 0.825 & 0.794 & 2.248 & 0.757 & 0.889 & 0.884 & 1.866 & 0.551
\tabularnewline

\hline
\end{tabular}}

\end{table*}

%% file: tables/table_seg_vs_reg.tex
\begin{table*}[!htb]
\centering
\caption{Table comparing the baseline RN (RUNet) with both the segmentation- and registration-based symmetry enforcement. Note that the registration-based refinement improves over the segmentation-based formulation, however, slightly increases the reconstruction time. The results are close to the best achievable ones (Table~\ref{tab:sn_limit}). Seg-SN and Reg-SN denote segmentation-based and registration-based symmetry enforcement, respectively.}
\renewcommand{\arraystretch}{1.0}
\footnotesize
\resizebox{0.99\textwidth}{!}{%
\begin{tabular}{lccccccccc}
\label{tab:seg_vs_reg}
Network & \multicolumn{4}{c}{SkullBreak} & \multicolumn{4}{c}{SkullFix} & \multicolumn{1}{c}{Avg. Time [s] $\downarrow$} \tabularnewline
\hline
\multicolumn{1}{c}{} & \multicolumn{1}{c}{DSC $\uparrow$} & \multicolumn{1}{c}{SDSC $\uparrow$}  & \multicolumn{1}{c}{HD95 [mm] $\downarrow$} & \multicolumn{1}{c}{MSD [mm] $\downarrow$} & \multicolumn{1}{c}{DSC $\uparrow$} & \multicolumn{1}{c}{SDSC $\uparrow$}  & \multicolumn{1}{c}{HD95 [mm] $\downarrow$} & \multicolumn{1}{c}{MSD [mm] $\downarrow$}
\tabularnewline
Baseline ($\alpha=0$) & 0.843 & 0.743 & 2.438 & 0.787 & 0.892 & 0.881 & 1.882 & 0.554 & 4.71
\tabularnewline
Seg-SN ($\alpha=1$) & 0.904 & 0.902 & 1.678 & 0.539 & 0.927 & 0.923 & 1.437 & 0.389 & 4.69
\tabularnewline
Reg-SN ($\lambda=1e5$)  & 0.931 & 0.926 & 1.379 & 0.381 & 0.941 & 0.945 & 1.342 & 0.338 & 14.31
\tabularnewline
Seg-SN ($\alpha=1$) + Reg-SN ($\lambda=1e5$)  & \textbf{0.936} & \textbf{0.932} & \textbf{1.312} & \textbf{0.372} & \textbf{0.945} & \textbf{0.951} & \textbf{1.278} & \textbf{0.334} & 14.34
\tabularnewline
\hline
\end{tabular}}

\end{table*}

%% file: tables/table_sota.tex
\begin{table*}[!htb]
\centering
\caption{Table comparing the proposed method to other state-of-the-art methods for SkullFix and SkullBreak datasets. Please note that not all papers report results for both SkullBreak and SkullFix datasets and some report results only to a limited number of digits.}
\renewcommand{\arraystretch}{1.0}
\footnotesize
\resizebox{0.99\textwidth}{!}{%
\begin{tabular}{lccccccc}
\label{tab:sota}
Method & \multicolumn{3}{c}{SkullBreak} & \multicolumn{3}{c}{SkullFix} \tabularnewline
\hline
\multicolumn{1}{c}{} & \multicolumn{1}{c}{DSC $\uparrow$} & \multicolumn{1}{c}{BDSC $\uparrow$} & \multicolumn{1}{c}{HD95 [mm] $\downarrow$} & \multicolumn{1}{c}{DSC $\uparrow$} & \multicolumn{1}{c}{BDSC $\uparrow$} & \multicolumn{1}{c}{HD95 [mm] $\downarrow$} \tabularnewline

Proposed (Segmentation-based) & 0.90 & 0.91 & 1.68 & 0.93 & 0.92 & 1.44
\tabularnewline
Proposed (Registration-based) & 0.94 & 0.94 & 1.31 & 0.95 & 0.95 & 1.28
\tabularnewline
\hline

Wodzinski et.al.~\cite{wodzinski2024improving} & 0.95 & 0.95 & 1.13 & 0.97 & 0.96 & 1.07
\tabularnewline

Mahdi et al.~\cite{mahdi2021u} & 0.8 & 0.81 & 3.42 & 0.9 & 0.93 & 3.59
\tabularnewline

Yang et al.~\cite{mahdi2021u} & 0.8 & 0.89 & 3.52 & N/A & N/A & N/A
\tabularnewline

Wodzinski et al.~\cite{wodzinski2021improving} & 0.9 & 0.93 & 1.60 & 0.93 & 0.95 & 1.48
\tabularnewline

Yu et al.~\cite{yu2021pca} & N/A & N/A & N/A & 0.8 & 0.77 & 3.68
\tabularnewline

Kroviakov et al.~\cite{kroviakov2021sparse} & N/A & N/A & N/A & 0.85 & 0.95 & 2.65
\tabularnewline

Pathak et al.~\cite{pathak2021cranial} & N/A & N/A & N/A & 0.9 & 0.95 & 2.02
\tabularnewline

Wodzinski et al.~\cite{wodzinski2023high} & 0.87 & 0.85 & 1.91 & 0.90 & 0.89 & 1.71
\tabularnewline

Friedrich et al.~\cite{friedrich2023point} & 0.87 & 0.89 & 2.45 & 0.90 & 0.93 & 1.69
\tabularnewline

Wodzinski et al.~\cite{wodzinski2022deep} & 0.89 & 0.93 & 1.60 & 0.93 & 0.95 & 1.47
\tabularnewline

Li et al.~\cite{li2023sparse} & N/A & N/A & N/A & 0.88 & 0.96 & 4.11
\tabularnewline

Mazzocchetti et al. et al.~\cite{mazzocchetti2024automatic} & 0.85 & 0.89 & 2.39 & 0.90 & 0.93 & 1.86
\tabularnewline

Kesornsri et al. et al.~\cite{kesornsri2024cranext} & 0.86 & N/A & 2.12 & N/A & N/A & N/A
\tabularnewline

\hline

\end{tabular}}

\end{table*}